\documentstyle[12pt]{article}
\advance\textheight by \topskip
\begin{document}
\vspace{-0.2cm}
\begin{flushright}
MSU--DTP--94/21\\  October 94
\vskip2mm
hep-th/9410217
\end{flushright}
\vskip2cm
\begin{center}
{\LARGE \bf INTEGRABLE SYSTEMS IN STRINGY GRAVITY\\}
\vskip1cm
{\bf D.V. Gal'tsov}
\vskip2mm
{\em Department of Theoretical Physics, Physics Faculty,\\
Lomonosov Moscow State University, 119899, Moscow, Russia.\\
e--mail: galtsov@grg.phys.msu.su}
\end{center}

\vskip2cm

\centerline{\bf Abstract}
\begin{quotation}
Static axisymmetric Einstein--Maxwell--Dilaton and stationary axisymmetric
Einstein--Maxwell--Dilaton--Axion theories in four space--time dimensions
are shown to be integrable by means of the inverse scattering transform method.
\vskip5mm
\noindent
PASC number(s): 97.60.Lf, 04.60.+n, 11.17.+y

\end{quotation}

\newpage

Both vacuum and electrovacuum Einstein equations enjoy a complete
integrability property being restricted to space--times admitting a
two--parameter Abelian group of isometries \cite{int}.
This entailes rich mathematical structures such as
an infinite set of non--local conservation laws \cite{cons} and
Backlund transformations \cite{back}. Similar integrability propertiy is
shared by more general gravity coupled systems
including scalar and vector fields, which follow  from certain
Kaluza--Klein (KK) models \cite{bgm}.
New generalizations of Einstein equations arise in the zero--slope limit
of the heterotic string theory, in four dimensions they include vector fields,
a dilaton, an axion, and moduli fields \cite{het}.
It was shown recently that a {\em pure} gravity
coupled to dilaton and axion is also  2--dim integrable \cite{bak}.
However, the most intriguing features of string--motivated gravity,
related to the black hole puzzle, are due to peculiar nature of the dilaton
coupling to {\em vector} fields \cite{ho}. Here we show that two stringy
gravity models including vector fields are also 2--dim integrable: static
axisymmetric Einstein--Maxwell--Dilaton (EMD) system with an arbitrary dilaton
coupling constant, and stationary axisymmetric
Einstein--Maxwell--Dilaton--Axion (EMDA) system.

Our reasoning is based on the sigma--model approach used earlier to prove
integrability of 2--dim reductions of vacuum and electrovacuum Einstein
equations \cite{sigma}.
It consists in a derivation of the 3--dim  sigma--model from 4--dim
theory in a space--time possessing a Killing vector field,
and a subsequent identification of the {\em symmetric space} structure
of the target space. This implies possibility of {\em zero--curvature}
representation of the equations of motion and applicability of the
inverse scattering transfrom method \cite{zah},
when the second Killing symmetry is imposed. The procedure
is rather well--known, so we just outline its main steps and fix our
notation.

Consider a general 4--dim coupled system of gravitational,
$U(1)$  vector, and some scalar massless fields.
Assuming the metric to admit a time--like Killing
symmetry, one can write the interval as
\begin{equation}
ds^2=-f(dt-\omega_idx^i)^2+f^{-1}h_{ij}dx^idx^j,
\end{equation}
where $f,\;\omega_i$, and the 3--metric $h_{ij}$
depend on the space coordinates $x^i,i=1, 2, 3,$ only. Then the $U(1)$ field
is  fully describable in terms of electric and magnetic potentials
 $v,\;a$. Usually, in conformity with the Einstein constraints,
a twist potential $\chi$
may be introduced to generate the rotation one--form $\omega_i$.
Together with $f$ and scalars, these variables may be
interpreted as a set of  scalar fields constituting a source for
$h_{ij}$. If there is no scalar potentials in the initial 4--dim
action, the theory will be equivalent to a 3--dim sigma model
\begin{equation}
S_\sigma=\frac{1}{2}\int \left({\cal R}-{\cal G}_{AB}(\varphi)\partial_i
\varphi^A \partial_j \varphi^B
h^{ij}\right)\sqrt{h}d^3x,
\end{equation}
where ${\cal R}$ is the 3--dim scalar curvature,
$\varphi^A=(f,\chi,v,a, {\rm scalar} {\rm fields}),\; A=1,...,K,$
and ${\cal G}_{AB}$ is the target space metric.

Suppose that the target space  is a {\em symmetric} riemannian space $G/H$
with $N$--parameter isometry group $G$ acting transitively on it ($H$ being
an isotropy subgroup), generated  by the set of $N$
Killing vectors forming the Lie algebra of $G,\;\;[{K_\mu, K_\nu }]
={C^\lambda}_{\mu\nu}K_\lambda, \;\; \mu,\nu,\lambda=1,...,N.$
Then the equations of motion for $\varphi^A$ will be equivalent to the set
of conservation laws  for Noether currents
\begin{equation}
\partial_i(h^{ij}\sqrt{h}J_i^\mu)=0,\quad
J_i^\mu=\tau_A^\mu \;\frac{\partial\varphi^A}{\partial x^i},
\end{equation}
built using the corresponding Killing one-forms
$
{\bf\tau}^\mu =\eta^{\mu\nu}K^A_\nu {\cal G}_{AB}d\varphi^B,
$
where  $\eta^{\mu\nu}$ is an inverse to the Killing--Cartan metric
$\eta_{\mu\nu}=k{C^\alpha}_{\mu\beta}{C^\beta}_{\nu\alpha}$.
With a proper choice of $k$ these one--forms
will satisfy Maurer--Cartan equation
\begin{equation}
d{\bf\tau}^\mu +\frac{1}{2}{C^\mu}_{\alpha\beta}{\bf\tau}^\alpha\wedge
{\bf\tau}^\beta =0.
\end{equation}

Let $e_{\mu}$ denote some matrix
representation of the Lie algebra of
$G,\; [e_\mu,e_\nu]={C^\lambda}_{\mu\nu}e_\lambda$.
Define the following matrix--valued connection one--form:
${\cal A}={\cal A}_Bd \varphi^B = e_\mu {\bf\tau}^\mu$.
In view of (4), the corresponding curvature vanishes,
\begin{equation}
{\cal F}_{BC}={\cal A}_{C,B}-{\cal A}_{B,C}+ [{\cal A}_B,{\cal A}_C]=0,
\end{equation}
and thus ${\cal A}_B$ is a pure gauge
\begin{equation}
{\cal A}_B= -(\partial_B g) g^{-1},\quad g\in G .
\end{equation}
The pull--back of ${\cal A}$
onto the configuration space ${x^i}$ is
equivalent to (3) and, hence, to the equations of
motion of the sigma--model. In terms of $g$ the Eqs. (3) read
\begin{equation}
d\{(\star dg)g^{-1}\}=0,
\end{equation}
where a star stands for a 3--dim Hodge dual.

Now impose an axial  symmetry condition, representing
the 3--metric in the Lewis--Papapetrou form:
\begin{equation}
h_{ij}dx^idx^j=e^{2\gamma}(d\rho^2 + dz^2)+\rho^2 d\varphi^2.
\end{equation}
Then (7) becomes equivalent to a modified chiral equation
\begin{equation}
(\rho g_{,\rho} g^{-1})_{,\rho} +  (\rho g_{,z} g^{-1})_{,z}=0,
\end{equation}
and the corresponding Lax pair with a complex
spectral parameter $\lambda$ can be found:
\begin{equation}
D_1\Psi=\frac{\rho U-\lambda V}{\rho^2+\lambda^2}\Psi,\quad
D_2\Psi=\frac{\rho V+\lambda U}{\rho^2+\lambda^2}\Psi.
\end{equation}
Here $V=\rho g_{,\rho}g^{-1},\; U=\rho g_{,z}g^{-1}, \; \Psi$
is a matrix "wave function", and
\begin{equation}
D_1=\partial_z-\frac{2\lambda^2}{\rho^2+\lambda^2}\partial_{\lambda},\quad
D_2=\partial_{\rho}+\frac{2\lambda \rho}{\rho^2+\lambda^2}\partial_{\lambda}
\end{equation}
are commuting operators; then (9) follows from the compatibility
condition $[D_1, D_2]\Psi=0$.
This linearization is sufficient to establish a desired integrability
property. An inverse scattering transform method \cite{zah}
can be directly applied to (10) to generate multisoliton
solutions, and an infinite--dimensional
algebra of a Geroch--Kinnersley--Chitre (GKC) type can be derived.

Let us apply this formalism to EMD and EMDA systems.
The first is described by the action
\begin{equation}
S=\frac{1}{16\pi}\int \left(R-2(\partial\phi )^2-
e^{-2\alpha\phi}F^2\right)\sqrt{-g}\;d^4x,
\end{equation}
where $\phi$ is the real scalar field (dilaton),
$F=dA$ is the Maxwell two--form,
$\alpha$ is the dilaton coupling constant.  For $\alpha =0$, (12)
reduces to the Brans--Dicke--Maxwell (BDM) action in the
Einstein frame (with the Brans--Dicke parameter $\omega =-1$).
For $\alpha =\sqrt{3}$,
(12) is derivable from the 5--dim KK--theory.

In conformity with the Maxwell equations following from (12), electric and
magnetic potentials can be introduced via
\begin{equation}
F_{i0}=\frac{1}{\sqrt{2}}\partial_iv,\quad
F^{ij}=-\frac{f}{\sqrt{2h}}e^{2\alpha\phi}\epsilon^{ijk}\partial_ka,
\end{equation}
while the twist potential $\chi$ is defined through
\begin{equation}
\tau_i=\partial_i\chi +v\partial_ia-a\partial_iv,\;
\tau^i=-f^2\epsilon^{ijk}\partial_j\omega_k/\sqrt{h},
\end{equation}
(3--dim indices are raised and lowered using $h_{ij}$).
The corresponding target space is five--dimensional ($K=5$), and
\begin{equation}
{\cal G}=\frac{1}{2f^2}\left(df^2+(d\chi+vda-adv)^2\right)+
\frac{1}{f}(e^{-2\alpha\phi}dv^2+e^{2\alpha\phi}da^2)+2d\phi^2.
\end{equation}
For $\alpha =0$ and $\phi = const$ this metric reduces to one given by
 Neugebauer and Kramer for the EM system  \cite{sigma}.

It turns out that, for the general stationary class of metrics (1), the target
space (15) is a symmetric riemannian space only for
${\alpha}=0,\sqrt{3}$, when it has the structure
of cosets $SU(2,1)/S(U(2)\times U(1)) \times R$ and $SL(3,R)/SO(3)$
respectively, corresponding to  BDM
and 5--dimensional KK theories. For $\alpha\neq 0,\sqrt{3}$ the
isometry group of (15) is only $N=5$ solvable subgroup of $SL(3, R)$.
However, if an additional condition of {\em staticity} is imposed,
$\omega_i=0$, the (truncated) target space possess the
symmetric space prperty for {\em arbitrary} $\alpha$.

In the static case it is consistent to consider
electric and magnetic configurations separately.
Both will be described by the same equations after reparametrization
\begin{equation}
\xi=(\alpha\phi-1/2\;\ln f)/\nu,\;\; \eta=(\phi+\alpha/2\; \ln f)/\nu,
\end{equation}
for a magnetic case, and
\begin{equation}
\xi=-(\alpha\phi+1/2\;\ln f)/\nu,\;\; \eta=(\phi-\alpha/2\; \ln f)/\nu,
\end{equation}
for an electric one, where $\nu=(\alpha^2+1)/2$.
Denoting as $u$ either magnetic ($a$) or electric ($v$)
potentials respectively, one can write the line element of the truncated
three--dimensional target space as
$dl_3^2=d\eta^2+dl_2^2$ where
\begin{equation}
dl_2^2=d\xi^2+e^{2\nu\xi} du^2
\end{equation}
Since $\eta$ decouples, it is sufficient to
deal only with this 2--dim space, which can easily be shown to
represent a coset $SL(2, R)/U(1)$. Indeed,
one can find three Killing vectors for (18):
\begin{equation}
K_1=\partial_u,\;\; K_2=p\partial_u-\nu^{-1}u \partial_{\xi},\;\;
K_3=u\partial_u-\nu^{-1}\partial_\xi,
\end{equation}
where $p=(u^2-\nu^{-2}e^{-2\nu\xi})/2$, with the $sl(2, R)$
structure constants ${C^3}_{12}={C^2}_{32}={C^1}_{13}=1$.
The corresponding Killing--Cartan
one--forms, with the normalization $k=(2\nu)^{-2}$, will satisfy (4), and
$dl^2_2=1/2\; \eta_{\mu\nu}{\bf\tau}^\mu\otimes{\bf\tau}^\nu$, where
$\eta_{\mu\nu}=2k \;{\rm diag} (1,1,-1)$.
Choosing as $e_\mu$ a $2\times 2$ representation of $sl(2, R)$,
one can find from (6) the following  matrix $g \in SL(2,R)/U(1)$:
\begin{equation}
g=\nu e^{\nu\xi}\sqrt{2}\left(\begin{array}{cc}
u^2-p & -u/\sqrt{2} \\
-u/\sqrt{2} & 1\\
\end{array} \right).
\end{equation}
Alternatively,  in view of the isomorphism $SL(2, R) \sim SO(2, 1)$,
a $3\times 3$ representation in terms of $SO(2, 1)/SO(2)$ coset can be
derived. In the axisymmetric case both can be used in the Lax pair (10).

For $\alpha=0$ ($\nu=1/2$) the above theory reduces to the corresponding
representation for electrovacuum. Since the underlying algebraic structure
is $\alpha$--independent, already this fact is sufficient to reveal
integrability of the static axisymmetric EMD system with arbitrary $\alpha$.
However, the  integrability of electrovacuum in
the {\em stationary} case is not shared by the arbitrary--$\alpha$ EMD system.

Remarkably, the EMDA theory turns out to be integrable
in the {\em stationary} axisymmetric case too.
The EMDA action in four dimensions reads
\begin{equation}
S=\frac{1}{16\pi}\int \left\{R-2\partial_\mu\phi\partial^\mu\phi -
\frac{1}{2} e^{4\phi}
{\partial_\mu}\kappa\partial^\mu\kappa
-e^{-2\phi}F_{\mu\nu}F^{\mu\nu}-\kappa F_{\mu\nu}{\tilde F}^{\mu\nu}\right\}
\sqrt{-g}d^4x,
\end{equation}
where ${\tilde F}^{\mu\nu}=\frac{1}{2}E^{\mu\nu\lambda\tau}F_{\lambda\tau},\;
\kappa$ is an axion field.
An electric potential is still introduced through the first
of Eqs.(13), while for $a$ one has
\begin{equation}
e^{-2\phi}F^{ij}+\kappa {\tilde F}^{ij}=-f\epsilon^{ijk}
\partial_ka/\sqrt{2h}.
\end{equation}
For a twist potential (14) remains valid. The target space now
is 6--dimensional ($K=6$), and its metric reads
\begin{equation}
{\cal G} =\frac{1}{2}e^{-4\phi}\omega_{\kappa}^2+2d\phi^2+
\frac{1}{2}\left( \frac{df^2}{f^2}+f^2\omega_{\chi}^2\right)
+f\left\{e^{2\phi}\omega^2_v+e^{-2\phi}\omega^2_a\right\},
\end{equation}
where
\[\omega_{\kappa}=e^{4\phi} d\kappa,\;\;\omega_{\chi}=f^{-2}(d\chi+vda-adv), \]
\begin{equation}
 \omega_v=f^{-1}e^{-2\phi} dv,\;
\omega_a=f^{-1}e^{2\phi} (da - \kappa dv).\;\;
\end{equation}
 Note, that the EMDA theory does not include the EMD one as a
particular case. Indeed, setting $\kappa=0$ gives a constraint
$F\tilde F=0$. Similarly, the EMD theory does not contain the EM one:
setting $\phi=0$ gives another constraint $F^2=0$.

As it was shown recently \cite{gk}, the space (23) possess a
$N=10$ isometry group consisting of scale, 3 gauge, 3
axion--dilaton duality, and 3 Ehlers--Harrison--type transformations,
which unify $T$ and $S$ string dualities in 4--dim zero--slope heterotic
string theory. Here we will show that the target space is a {\em symmetric}
space which can be identified with the coset $SO(3,2)/(SO(3)\times SO(2))$.
Denoting generators of $SO(3,2)$  by pair indices $ab, a<b$, where
$a,b=0,\theta, 1,2,3$ correspond to  the invariant
 metric $G_{ab}={\rm diag}(-1,-1,1,1,1)$, one has
\begin{equation}
\left[M_{ab} M_{cd}\right]=G_{bc}M_{ad}-G_{ac}M_{bd}+G_{ad}M_{bc}-G_{bd}M_{ac}.
\end{equation}
The set of 10 one--form satisfying Maurer--Cartan equations with the
structure constants ${C^{cd}}_{ab\;ef}$ from (25) reads as follows.
An abelian subalgebra of $so(3,2)$ corresponds to
\begin{equation}
-\tau^{01}=\omega_1 +\omega_f,\;\; \tau^{\theta 2}=\omega_1
-\omega_f-2\omega_2,
\end{equation}
where
\begin{equation}
\omega_1=\kappa \omega_{\kappa}-2d\phi+a(v\omega_{\chi}+2\omega_a),\;
\omega_f=f^{-1}df+\chi \omega_{\chi},\;\; \omega_2=v\omega_v+\tilde a \omega_a.
\end{equation}
Introduce a reccurent sequence
\[\omega_3=\kappa \omega_a - \omega_v,\;
\omega_4=a\omega_{\chi} + \omega_3,\;
\omega_5=v\omega_{\chi} + \omega_a, \]
\[ \omega_6=d\kappa -\kappa^2\omega_{\kappa} +4\kappa d\phi -
a(\omega_4 + \omega_3),\;\;
\omega_7=\omega_{\kappa} +v(\omega_a + \omega_5),\]
\begin{equation}
\omega_8=a\tau^{01} -v\omega_6-\chi \omega_3 +a\omega_2+da,\;\;
\end{equation}
\[\omega_9=v\tau^{\theta 2} -a\omega_7-\chi \omega_a +v\omega_2+dv,\]
\[\omega=a\omega_9 -v\omega_8+\chi(\chi\omega_{\chi}-\omega_2-2\omega_f)
+d\chi. \]
Then the remaining set will read
\[2\tau^{0\theta}=\omega+\omega_6-\omega_7-\omega_{\chi},\;\;
2\tau^{02}=\omega-\omega_6-\omega_7+\omega_{\chi},\]
\begin{equation}
-2\tau^{\theta 1}=\omega+\omega_6+\omega_7+\omega_{\chi},\;\;
2\tau^{12}=\omega-\omega_6+\omega_7-\omega_{\chi},
\end{equation}
\[-\tau^{03}=\omega_5+\omega_8,\; \tau^{13}=\omega_5-\omega_8,\;
\tau^{\theta 3}=\omega_4-\omega_9,\; -\tau^{23}=\omega_4+\omega_9.\]
In terms of $\tau^{ab}$ one has ${\cal G}_{AB}=
1/2\; \eta_{ab\;cd} \tau^{ab}_A \tau^{cd}_B$, where $\eta_{ab\;cd}=
1/12\; {C^{gh}}_{ab\;ef} {C^{ef}}_{cd\;gh}$.

Now, using an adjoint representation of $so(3,2)$, one can build
$5\times 5$ connection one--form ${\cal A}$ and the corresponding matrix
$g \in SO(3,2)/(SO(3)\times SO(2))$. Fortunately,
due to isomorphism $SO(3,2)\sim Sp(2,R)$,
there exists  also more concise representation in terms of $4\times 4$
matrices. The symplectic connection reads
\begin{equation}
{\cal A}=\left(\begin{array}{cc}
C & D \\
F & -C^T \\
\end{array} \right),
\end{equation}
where $D, F, C$ are $2\times 2$ matrices, $D^T=D,\;\; F^T=F$,
\[ C=\frac{1}{2}\left\{\tau^{03}I_2 -\tau^{\theta 2}\sigma_x-
i\tau^{12}\sigma_y+\tau^{\theta 1}\sigma_z\right\},\]
\begin{equation}
 D=\frac{1}{2}\left\{ (\tau^{0
\theta} - \tau^{\theta 3})I_2 +
 (\tau^{23} - \tau^{02})\sigma_x +
 (\tau^{01} - \tau^{13})\sigma_z \right\},
\end{equation}
\[  F=\frac{1}{2}\left\{ -(\tau^{0\theta} + \tau^{\theta 3})I_2 -
 (\tau^{23} + \tau^{02})\sigma_x +
 (\tau^{01} + \tau^{1 3})\sigma_z \right\}.\]
Here $I_2$ is a unit matrix and $\sigma_x, \sigma_y, \sigma_z$ are Pauli
matrices with $\sigma_z$ diagonal. In view of (4), the equations of motion
of the EMDA sigma--model are equivalent to vanishing of the curvature
(5) related to (30). This implies the existence of the symmetric symplectic
$4\times 4$ matrix $g \in Sp(2, R)/U(2)$ entering Belinskii--Zakharov
representation.

To summarize: we have shown that target spaces corresponding to the
static EMD with an arbitrary dilaton coupling and the stationary EMDA
systems in 4  dimensions are symmetric Riemannian spaces isomorphic
to cosets $SO(2,1)/SO(2)$ and $SO(3,2)/(SO(3)\times SO(2))$ respectively
(or, equivalently, $SL(2,R)/U(1)$ resp. $Sp(2,R)/U(2)$). This ensures
zero--curvature representation of the equations of motion and
existence of the Lax--pair in the axisymmetric case.
The inverse scattering transform
method can be applied to deal with both systems, in particular,
to construct multisoliton solutions.  Current
algebras associated with $SL(2,R)$ and $Sp(2,R)$ generate
infinite--dimensional GKC--type symmetries.
Obviously, the whole reasoning
can be generalized to the case of a space--like initial Killing vector field,
as well as to the case of the Euclidean signature of the 4--space.

As it was noted in \cite{gk}, the isometry group of the EMDA
target space is larger than the product  of well--known
$T$ and $S$ string dualities \cite{dual}. Now it is clear that,
on the class of space--times admitting a two--parameter Abelian isometry
group, both these symmetries are particular elements of the
infinite--dimensional GKC--type group. The implications of this
to the {\em exact} string theory are still to be explored.
An intriguing question is whether classical integrability of the
2--dim reduced EMDA system entails the possibility of
explicit construction of new classes of exact string backgrounds
in terms of the gauged WZW models. This issue will be discussed
in a subsequent publication.

The work  was  supported by the Russian Foundation for Fundamental Research
grant 93--02--16977,  the ISF grant M79000, and the INTAS grant 93-3262.

\end{document}